\documentclass[journal,twocolumn,letterpaper]{IEEEJERM}
\renewcommand{\IEEEkeywordsname}{Index Terms}

\ifCLASSINFOpdf
\else
\fi

\usepackage{times,amsmath,epsfig}
\usepackage{fancyhdr}
\usepackage{amsmath}
\usepackage{amsfonts}
\usepackage{amssymb}
\usepackage[latin1]{inputenc}
\usepackage{array}
\usepackage{graphicx}
\usepackage{url}
\usepackage{subfigure}
\usepackage{bm}
\usepackage{breqn}
\usepackage{xcolor}
\usepackage{soul}
\usepackage{amssymb}
\usepackage{epstopdf}
\usepackage{orcidlink}
\usepackage{xcolor}
\begin{document}

%
\title{A Lumped RC Equivalent Circuit of Head Tissues for Dispersive Neuro-Electromagnetic Modeling}

\author{%
Angelo~Faccia\,\orcidlink{0009-0006-7879-4252},~\IEEEmembership{Graduate~Student~Member,~IEEE}, 
Ermanno~Citraro\,\orcidlink{0009-0005-6157-5902}, and~Francesco~P.~Andriulli\,\orcidlink{0000-0001-8359-7348},~\IEEEmembership{Fellow,~IEEE}%
}

\markboth{IEEE Journal of Electromagnetics, RF and Microwaves in Medicine and Biology}
{Z. Peng \MakeLowercase{\textit{et al.}}: Demo of IEEE Journal of Electromagnetics, RF and Microwaves in Medicine and Biology (JERM)}

\twocolumn[
\begin{@twocolumnfalse}
  
\maketitle

\begin{abstract}
Accurate modeling of electric potential and current distribution in head tissues is crucial for the design and evaluation of neuro-sensing and neuro-stimulation systems operating in the sub-megahertz frequency range.
Numerical methods are widely employed in electromagnetic simulations, however their computational cost can limit their applicability to rapid prototyping, real-time simulations, and circuit-level integration.

In this work, we introduce a lumped RC equivalent circuit model that reproduces the electrical behavior of a canonical three-layer spherical head geometry over a frequency range up to 50~kHz. The model accounts for frequency-dependent tissue conductivity and permittivity to capture dispersive effects, employing complex conductivity in the electro-quasi-static (EQS) regime. The circuit topology uses a minimal set of impedance elements in order to represent the essential mechanisms of electric signal propagation.

Validation was performed using a dipolar brain source configuration for scalp voltage peak estimation, showing close agreement with semi-analytical solutions across different skull thicknesses and dipole eccentricities. In addition, the impact of tissue dispersion and capacitive branches on the model predictions was quantitatively assessed, showing their contribution to the overall fidelity of the proposed approach. 
\end{abstract}



\end{@twocolumnfalse}]

\begingroup
\renewcommand{\thefootnote}{}%
\renewcommand{\footnoterule}{}%
\footnotetext{%
\hspace{0pt}\raggedright
The authors are with the Department of Electronics and Telecommunications (DET), 
Politecnico di Torino, 10129 Turin, Italy (e-mail: angelo.faccia@polito.it; 
ermanno.citraro@polito.it; francesco.andriulli@polito.it).%
}
\endgroup

%
\IEEEpeerreviewmaketitle

\section{Introduction}
%
%
%
%

\IEEEPARstart{T}{he} Measurement of brain electrical activity through scalp electrodes, known as electroencephalography (EEG), provides a noninvasive means of monitoring neuronal activity with high temporal resolution~\cite{nunez2006electric,berger1929human}. Modeling the scalp potentials generated by neural sources, commonly represented as current dipoles~\cite{de1988mathematical}, is known as the EEG forward problem~\cite{hallez2007review}. This modeling step is fundamental for solving the corresponding inverse problem, which aims to reconstruct the spatial distribution of neural activity from the measured scalp signals~\cite{grech2008review}. This procedure plays a crucial role in pre-surgical screening in drug-resistant epilepsy and real-time Brain-Computer Interface (BCI) systems~\cite{van2020ictal,kamousi2005eeg,kaiboriboon2012eeg}.

Accurate representation of signal propagation in head tissues is not only important for EEG-related techniques but also for numerous biomedical applications, ranging from noninvasive sensing to targeted stimulation. It is increasingly relevant for understanding how biological tissues interact with electromagnetic fields over different frequency bands, ensuring both precision and safety across a wide range of applications~\cite{adey1981tissue,dornhof2019electrical}. These techniques encompass transcranial direct and alternating current stimulation (tDCS/tACS) in the sub-100~Hz range~\cite{paulus2011transcranial}, low-, high-, and ultra-high-frequency deep brain stimulation (LF/HF/UHF-DBS) protocols used in Parkinson's disease treatment, which can work up to a few kilohertz~\cite{harmsen2019p,benabid2003deep}, and electrosurgical procedures that typically operate around 1~MHz~\cite{dornhof2019electrical,malis1996electrosurgery}.

The safety and precision required by these applications highlight the need for solutions capable of accurately modeling the electrical behavior of head tissues across a broad frequency spectrum. Such broadband consistency requires an accurate representation of biological tissue dispersion as the operating frequency extends beyond the range typically associated with endogenous brain activity~\cite{wagner2014impact,gabriel1996dielectric}.

At the same time, the growing adoption of implantable and wearable biomedical electronic systems for real-time neuromodulation underscores the importance of compact equivalent-circuit models of the head, designed to reproduce the electrical behavior of head tissues with minimal computational cost~\cite{chen2018multi,shon2017implantable}. The integration of these models into circuit-simulation environments would enable rapid design and testing, while their deployment in real-time controllers would support adaptive, closed-loop tuning of stimulation parameters.

Numerical techniques such as the Finite Element Method (FEM), Finite Difference Method (FDM), and Boundary Element Method (BEM)~\cite{jin2015theory,hallez2007review,salvador2010modeling} are commonly used to compute the electric field distribution within head tissues, as they can account for complex geometries and heterogeneous properties. However, their computational requirements may pose challenges for their direct integration within circuit simulation environments and real-time applications.

Partial solutions have been explored with circuit-based formulations such as the Resistor Mesh Model (RMM), ~\cite{franceries2003solution,chauveau2004effects}, where the head volume is discretized into a network of resistors connecting spherical or cubic voxels.
This approach models static head behavior; however, because it relies on a fine volumetric discretization rather than a compact lumped equivalent circuit, it entails non-negligible computational demands and does not aim to reproduce the capacitive properties of biological tissues.
Alternatively, compact equivalent-circuit models are motivated by cell- and membrane-based descriptions of tissue electrical behavior~\cite{mcadams1995tissue}.
These models have been used to characterize the electrical response of individual tissues in different applications~\cite{coston2003transdermal,dornhof2019electrical}.
In contrast, limited work has addressed the development of lumped element equivalent circuit representations of the entire human head that remain accurate over a broad frequency range, while accounting for tissue dispersion and displacement-current effects. 
Therefore, a solution that does not suffer from these limitations, combining compactness, circuit compatibility, and consistency with the frequency-dependent electrical behavior of biological tissues, is still desirable.

In this work, we propose a lumped-element circuit model derived from the canonical three-layer spherical head geometry, representing the scalp, skull, and brain compartments. The model addresses the quasi-static volume-conduction problem linking internal current sources to peak scalp potentials, enabling application to both neural sensing and stimulation via the reciprocity theorem~\cite{welch2003reciprocity,rush2008eeg}, which provides a link between forward modeling and brain neuromodulation~\cite{fernandez2016transcranial,dmochowski2017optimal}.
The circuit topology is composed of a compact set of impedance elements defined by the geometrical configuration and electrical properties of the tissue layers. The model not only includes explicit capacitive branches to account for displacement currents but also incorporates the frequency dispersion of both electrical conductivity and permittivity.

The proposed circuit model was first validated against a semi-analytical reference formulation based on scalar spherical harmonics expansion by comparing the peak scalp potential generated by a radially oriented current dipole source~\cite{arthur1970effect}. Following validation, numerical analyses were performed to quantify the impact of displacement currents and frequency dispersion on circuit voltage estimation. The results indicated that these effects produce non-negligible variations in potential amplitude, thereby supporting their inclusion in the equivalent circuit model to ensure a physically consistent electrical representation of head media.

This paper is organized as follows. Section \ref{sec:theo_back} presents background material and introduces the notation. Section \ref{sec:methods} describes the proposed circuit equivalent model, detailing its architecture and parameter dependencies. Section~\ref{sec:results} presents the validation of the proposed circuit model against the semi-analytical formulation and quantifies the errors introduced when neglecting capacitive or dispersive effects. Section~\ref{sec:discussion} discusses the results and outlines future research directions. Preliminary and partial
results have been presented in the conference contribution~\cite{facciaAPS2026}.

\begin{figure}
    \centering
    \includegraphics[width=1\linewidth]{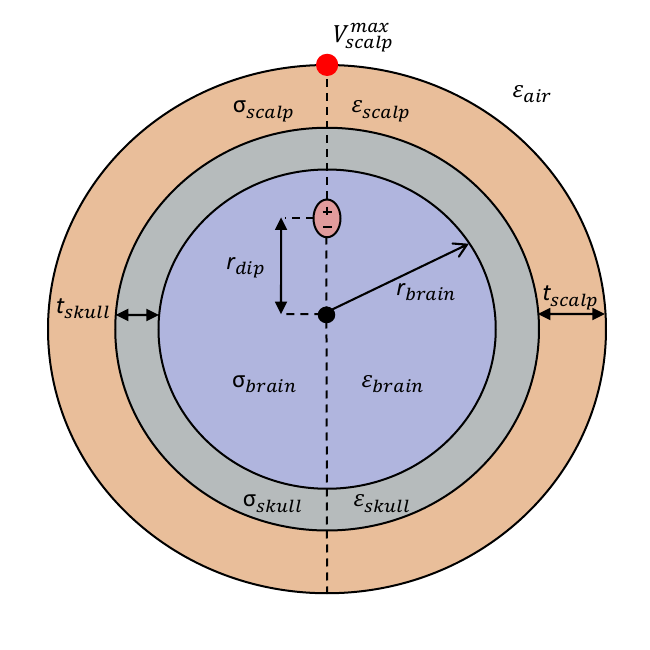}
    \caption{
Three-layer spherical head model (with an additional infinite air layer) composed of scalp, skull, and brain compartments. 
A radial current dipole is located within the brain at a distance \(r_{\mathrm{dip}}\) from the center. 
Each tissue layer is characterized by frequency-dependent conductivity \(\sigma(\omega)\) and permittivity \(\varepsilon(\omega)\). 
Standard geometrical parameters are set as: brain radius \(r_{\mathrm{brain}} = 7.91~\mathrm{cm}\), skull thickness \(t_{\mathrm{skull}} = 5.9~\mathrm{mm}\), and scalp thickness \(t_{\mathrm{scalp}} = 7~\mathrm{mm}\).
}
    \label{fig:headmodel}
\end{figure}
\section{Theoretical Background}
\label{sec:theo_back}
Under the quasi-static approximation, commonly employed below the MHz range where inductive effects are negligible~\cite{plonsey1967considerations}, the electric field is irrotational and admits a scalar potential, $\mathbf{E} = -\nabla V$. The generalized Poisson equation reads:
\begin{equation}
\nabla \cdot \left( \sigma^c(\omega) \nabla V \right) = \nabla \cdot \mathbf{J}_i,
\end{equation}
\cite{hallez2007review}, where $\mathbf{J}_i$ denotes the impressed current density and $\sigma^c(\omega) = \sigma(\omega) + i \omega \varepsilon(\omega)$ represents the complex frequency-dependent conductivity under time-harmonic conditions, encapsulating both conductive and capacitive responses of biological tissues.

Along with Poisson's equation, Dirichlet and Neumann boundary conditions ensure the physical continuity between adjacent head compartments for both the scalar potential $V$ and the normal component of the total current density, which includes conductive and displacement contributions~\cite{hallez2007review,kybic_common_2005}:
\begin{equation}
V_- = V_+, \qquad
(\sigma_- + i \omega \varepsilon_-)\, E^n_-
=
(\sigma_+ + i \omega \varepsilon_+)\, E^n_+,
\end{equation}
where $E^n = \hat{\mathbf{n}} \cdot \mathbf{E} = -\frac{\partial V}{\partial n}$ is the normal component of the electric field at the interface, with $\hat{\mathbf{n}}$ pointing outward from inner domain $-$ to outer domain $+$.


Both the electrical conductivity $\sigma(\omega)$ and the permittivity $\varepsilon(\omega)$ of biological tissues exhibit frequency-dependent behavior due to dispersive effects, which have been extensively characterized in the literature through \textit{in vitro} and \textit{in vivo} measurements~\cite{gabriel1996dielectric,wagner2014impact}. 

Simplified head models, such as the piecewise-homogeneous, concentric, three-shell spherical geometry illustrated in Fig.~\ref{fig:headmodel}, allow semi-analytical solutions of Poisson's equation via scalar spherical harmonics (SSH) expansion~\cite{arthur1970effect,salu1990improved,jackson1999classical}. The formulation is extended to the general conduction regime, including both displacement current contributions and tissue dispersive effects.

Let \(\sigma^c_i\) denote the complex conductivity of the \(i\)-th layer, and define the geometrical ratios as \(\eta = \frac{r_{\text{dip}}}{r_1}\) and \(\psi_{ij} = \frac{r_i}{r_j}\), where \(r_{\text{dip}}\) is the dipole radial position and \(r_i\) represents the radius of the interface between layers \(i\) and \(i+1\). The scalp potential peak generated by a radial brain current dipole with a dipole moment \(p_r\) in a spherical head model composed of the three tissue layers and an external air region can be expressed as
\begin{equation}
V_{SSH}^{\mathrm{max}}(\omega, r_{3}) = \frac{p_r(\omega)}{4\pi \sigma^c_{4}(\omega) r_{3}^2} 
\sum_{l=1}^{\infty} A_{4}(l,\omega),
\label{eq:anal}
\end{equation}
where layers are indexed as \(1\) for the brain, \(2\) for the skull, \(3\) for the scalp, and \(4\) for the surrounding air.

The coefficient \(A_{4}(l,\omega)\) is given by
\begin{equation}
A_{4}(l,\omega)= 
\frac{l(2l+1)^3\,\sigma^c_{2}\sigma^c_{3}\sigma^c_{4}\,\eta^{l-1}\psi_{13}^{l-1}}
{l(l+1)(\psi_{23}^{2l+1}X_{1} + \psi_{13}^{2l+1}X_{2} + \psi_{12}^{2l+1}X_{3}) + \tilde{X}},
\end{equation}
where the auxiliary terms are defined as
\[
X_1 = \tilde{\sigma}^c_{21} \sigma^c_{32} \sigma^c_{43}, \quad
X_2 = \sigma^c_{21} \tilde{\sigma}^c_{23} \sigma^c_{43}, \quad
X_3 = \sigma^c_{21} \sigma^c_{32} \tilde{\sigma}^c_{43},
\]
\[
\tilde{X} = \tilde{\sigma}^c_{21} \tilde{\sigma}^c_{32} \tilde{\sigma}^c_{43},
\]
with \(\sigma^c_{ij} = \sigma^c_i - \sigma^c_j\) and \(\tilde{\sigma}^c_{ij} = (l+1)\sigma^c_i + l\sigma^c_j\).

Although this semi-analytical approach offers a rigorous benchmark, its confinement to spherical geometries and the lack of an electrical analogue motivate the design of a compact surrogate circuit model.
\section{Methods and Procedures}
\label{sec:methods}
The circuit model developed in this work reproduces the canonical, piecewise homogeneous, three-shell spherical head structure (brain, skull, scalp, and surrounding air) through an equivalent lumped electrical network. It is designed to capture the dominant mechanisms of bioelectric field propagation while retaining validity over the $10~\mathrm{Hz}$--$50~\mathrm{kHz}$ frequency range. By replacing full volumetric methods with a lightweight circuit representation, the model aims to mitigate numerical burdens. As an illustrative application, the model is employed to compute the peak scalp voltage generated by a radially oriented brain source, showing consistency with the semi-analytical spherical-harmonics solution in \eqref{eq:anal} and providing a foundation for circuit equivalents of more complex and realistic geometries.
\begin{figure}[t]
    \centering
    \includegraphics[width=1\linewidth]{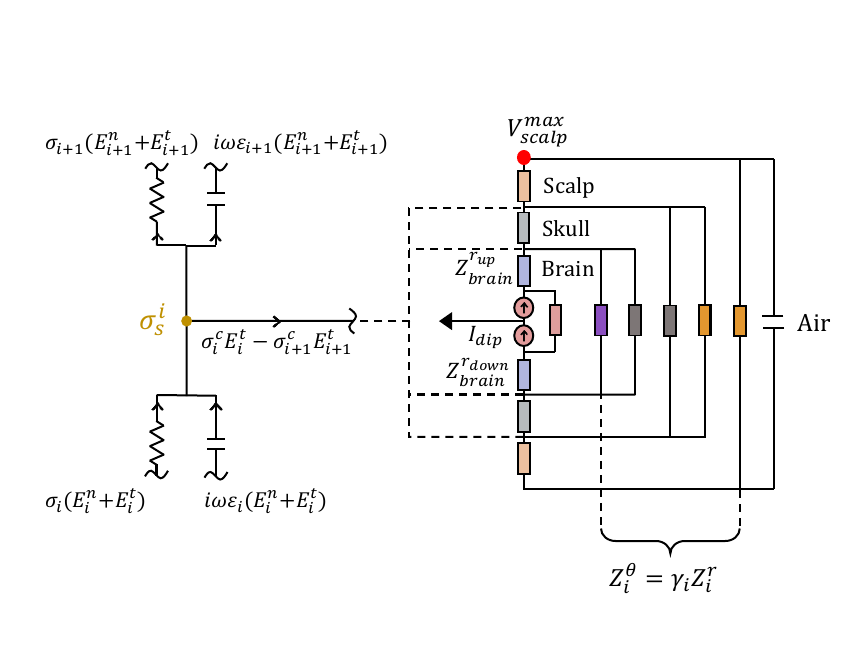}
    \caption{Lumped-circuit model of a three-layer head with a radially oriented dipolar brain source. Vertically aligned branches implement the radial impedances $Z_i^{r}$ of each tissue layer; horizontally aligned branches implement the tangential impedances $Z_i^{\theta}$. Kirchhoff's current law (KCL) at every node enforces the Neumann boundary condition between adjacent layers $i$ and $i+1$, including the effect of the interfacial surface charge density $\sigma^i_{s}$ that arises from permittivity discontinuities. The brain radial impedances $Z_{\text{brain}}^{r_{\text{up}}}$ and $Z_{\text{brain}}^{r_{\text{down}}}$ differ according to dipole eccentricity. The source $I_{\text{dip}}$ models the current generated by dipolar neuronal activity and is related to the current dipole moment $p_r$ and the effective dipole length $d$ by $I_{\text{dip}} = p_r/d$. The point $V_{\text{scalp}}^{\max}$ denotes the maximum scalp voltage.}
    \label{fig:circuit_model}
\end{figure}

\subsection{Topological Structure}

Fig.~\ref{fig:circuit_model} illustrates the overall circuit topology. 
The network was structured to employ a minimal number of impedance elements while preserving the dominant mechanisms governing current propagation through the head tissues. 
Each tissue compartment was treated as a homogeneous subdomain, whose electrical response depends jointly on its intrinsic material properties and on its geometrical configuration within the concentric structure of the head. 
The tissue properties were embedded in the circuit by defining each resistance as inversely proportional to the corresponding layer conductivity and each capacitance as directly proportional to the layer absolute permittivity, whereas the geometric factors $\Gamma_i$ encoded the spatial dependence on the layer interface radius $r_i$ and thickness $t_i$:
\[
R_i = \frac{\Gamma{(r_i,t_i)}}{\sigma_i(\omega)}, \qquad 
C_i = \Gamma{(r_i,t_i)} \varepsilon_i(\omega).
\]

A straightforward radial stacking of impedances would mimic the layered anatomy but would fail to capture the correct distribution of current within each tissue. 
In reality, not all electric-field lines generated by a dipolar source traverse the entire volume: part of the current crosses successive layers, while another part propagates within the same shell before returning back to the dipole.
To account for both mechanisms, each layer \(i\) was therefore represented by two coupled impedance branches: 
a \emph{radial} path, \(Z_i^{r}\), describing through-layer current flow, and a \emph{tangential} path, \(Z_i^{\theta}\), representing intra-layer return currents.
Each branch is implemented as a parallel RC element. 

Model fidelity to the underlying electromagnetic behavior was enforced via Kirchhoff's Current Law (KCL) at each interfacial node, consistent with the Neumann boundary condition on the total current density between layers $i$ and $i+1$ at the interface $r_i$, as illustrated in Fig.~\ref{fig:circuit_model}:
\begin{equation}
(\sigma_i + i \omega \varepsilon_i)\, E^n_i(r_i)
=
(\sigma_{i+1} + i \omega \varepsilon_{i+1})\, E^n_{i+1}(r_i).
\end{equation}
This ensures normal current continuity across interfaces and naturally reproduces the surface-charge accumulation \(\sigma_s^i\) associated with permittivity discontinuities, which is stored in the radial nodes:
\begin{equation}
\sigma^i_{s} = \left( \varepsilon_{i+1} {E}_{i+1}^n - \varepsilon_i {E}_i^n\right).
\end{equation}
The remaining current closes within each subdomain via the tangential impedance branches \(Z_i^{\theta}\).

The relative strength of radial and tangential conduction paths depends on the aspect ratio of each tissue layer. 
To capture this dependence, a dimensionless geometric coefficient $\gamma_i$ was introduced to scale the tangential branch of layer \(i\) relative to its radial counterpart:
\[
\gamma_i = \frac{Z_i^{r}}{Z_i^{\theta}} 
= \frac{R_i^{r}+i\omega C_i^{r}}{R_i^{\theta}+i\omega C_i^{\theta}}
= \frac{\Gamma_{(r_i,t_i)}^{r}}{\Gamma_{(r_i,t_i)}^{\theta}}.
\]
This factor depends solely on the radii and thicknesses of the spherical shells and is independent of any material anisotropy, providing a purely geometrical weighting between the two conduction modes.

Finally, the position of the current dipole, assumed to be located within the brain compartment, introduces asymmetry in the distribution of brain current. 
This effect was modeled by splitting the brain's radial impedance into two branches, \(Z_{\mathrm{brain}}^{r_{\mathrm{up}}}\) and \(Z_{\mathrm{brain}}^{r_{\mathrm{down}}}\), whose relative values depend on the dipole eccentricity parameter \(\eta\), shifting the source position along the radial axis.

\subsection{Geometrical Variability Parametrization}

The geometrical parameters \(\gamma_i\) were first calibrated under a reference configuration defined by a canonical three-layer spherical geometry 
(brain radius $r_{\mathrm{brain}} = 7.91\,\mathrm{cm}$, skull radius $r_{\mathrm{skull}} = 8.50\,\mathrm{cm}$, and scalp radius $r_{\mathrm{scalp}} = 9.20\,\mathrm{cm}$, as shown in Fig.~\ref{fig:headmodel}), 
with a dipole placed at the center ($\eta = 0$) and non-dispersive static electrical tissue properties. 
The dipole moment was set to $p_r = 15~\mathrm{nA{\cdot}m}$, corresponding to a typical pyramidal neuronal group of size $d = 1~\mathrm{mm}$, 
yielding an equivalent dipolar current $I_{\mathrm{dip}} = p_r/d$.

This calibration step established the circuit response for the standard head geometry through numerical optimization, 
so that the resulting peak scalp potential matches the semi-analytical spherical-harmonics formulation, under the assumption that $\gamma_i$ depends only on the geometrical configuration (radii and thicknesses) and not on the electrical properties of the tissues.

Subsequently, the model was parametrically extended to account for varying dipole eccentricity ($\eta$) and the geometric variability of the tissue layers.
\vspace{3pt}
\paragraph*{\textbf{Dipole eccentricity} $\bm{\eta}$}  
Neural sources detected by brain sensing techniques, as well as stimulation targets in tDCS and tACS, are typically located within cortical  regions. 
It is therefore essential that the proposed model maintains accuracy for highly eccentric dipole positions, approaching the brain-skull interface.

\begin{figure}
    \centering
    \includegraphics[width=1\linewidth]{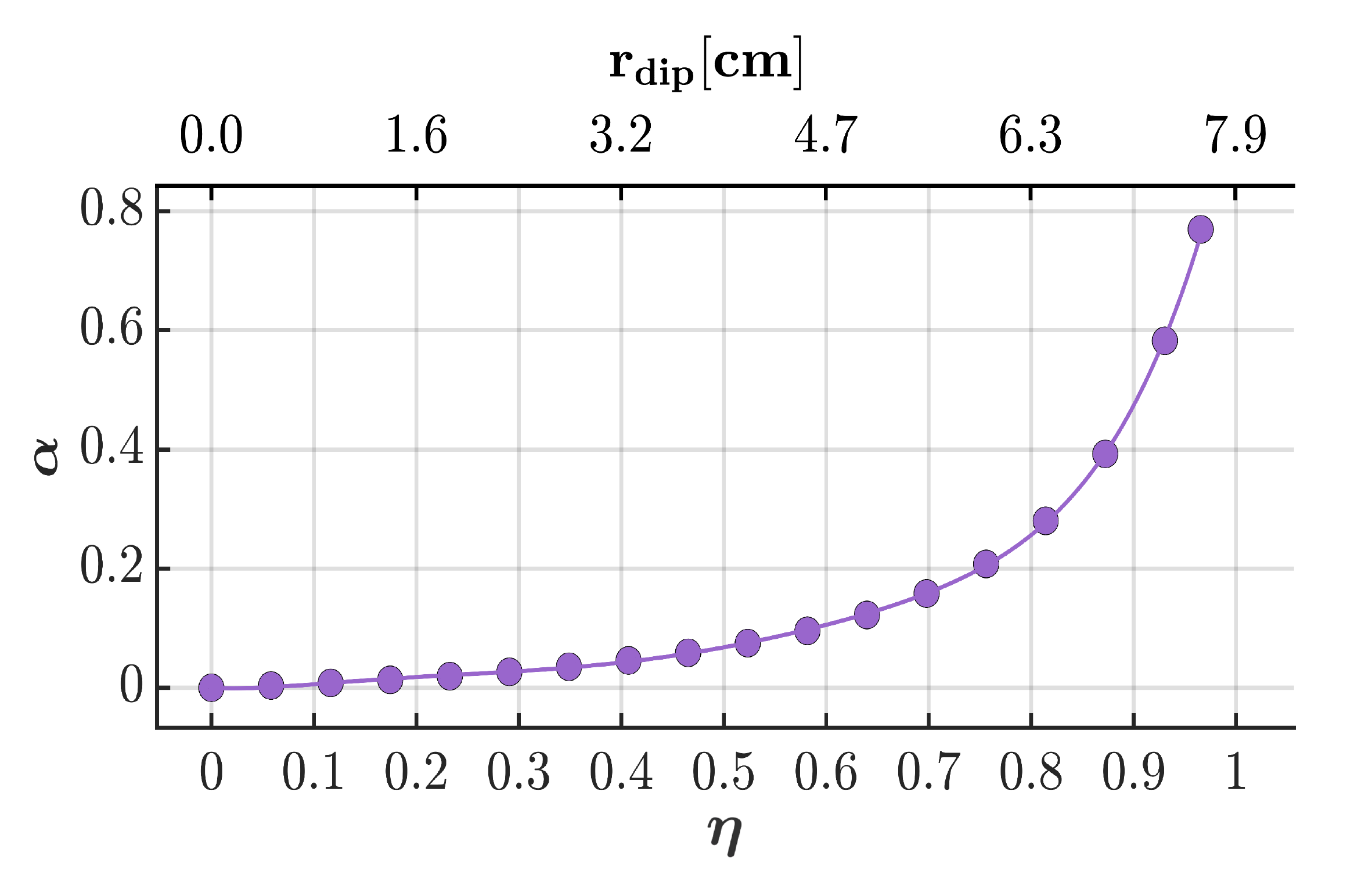}
    \caption{Polynomial fit of the $\alpha$ parameter as a function of dipole eccentricity, \(\eta\) (bottom axis) and dipole radius, \(r_{dip}\) (upper axis). The best-fit curve is a six-order polynomial (RMSE = $2.33\times10^{-3}$).}
    \label{fig:eccfit}
\end{figure}

To account for this effect, the brain radial impedance was divided into two branches, $Z^{r_{\mathrm{up}}}_{\mathrm{brain}}$ and $Z^{r_{\mathrm{down}}}_{\mathrm{brain}}$, representing the asymmetric current paths produced by an off-center source.  
An asymmetry parameter $\alpha(\eta)$ was introduced to redistribute the total brain radial impedance $Z_{\mathrm{brain}}^0$ between the two sub-blocks:
\[
Z^{r_{\mathrm{up}}}_{\mathrm{brain}} = Z_{\mathrm{brain}}^0 \left[1 - \alpha(\eta)\right], \quad
Z^{r_{\mathrm{down}}}_{\mathrm{brain}} = Z_{\mathrm{brain}}^0 \left[1 + \alpha(\eta)\right].
\]
The parameter $\alpha(\eta)$ was obtained through a polynomial interpolation of numerically optimized data based on the semi-analytical reference solution, over the range $\eta \in [0,\,0.966]$, corresponding to dipoles located as close as 2.7\,mm from the brain-skull interface, as illustrated in Fig.~\ref{fig:eccfit}.
\vspace{3pt}
\paragraph*{\textbf{Layers' Dimension}}
Considering the variability in head tissue morphology reported across subjects~\cite{antonakakis2020inter}, the model was designed to maintain accuracy under physiological variations of the interface radii. This was achieved by establishing simple relationships between the layer-specific coefficients ($\gamma_i$) and the corresponding layer geometries, thereby modeling the current partitioning between radial and tangential pathways. To characterize these dependencies, three independent parameter sweeps were performed.

\begin{figure}
    \centering
    \includegraphics[width=1\linewidth]{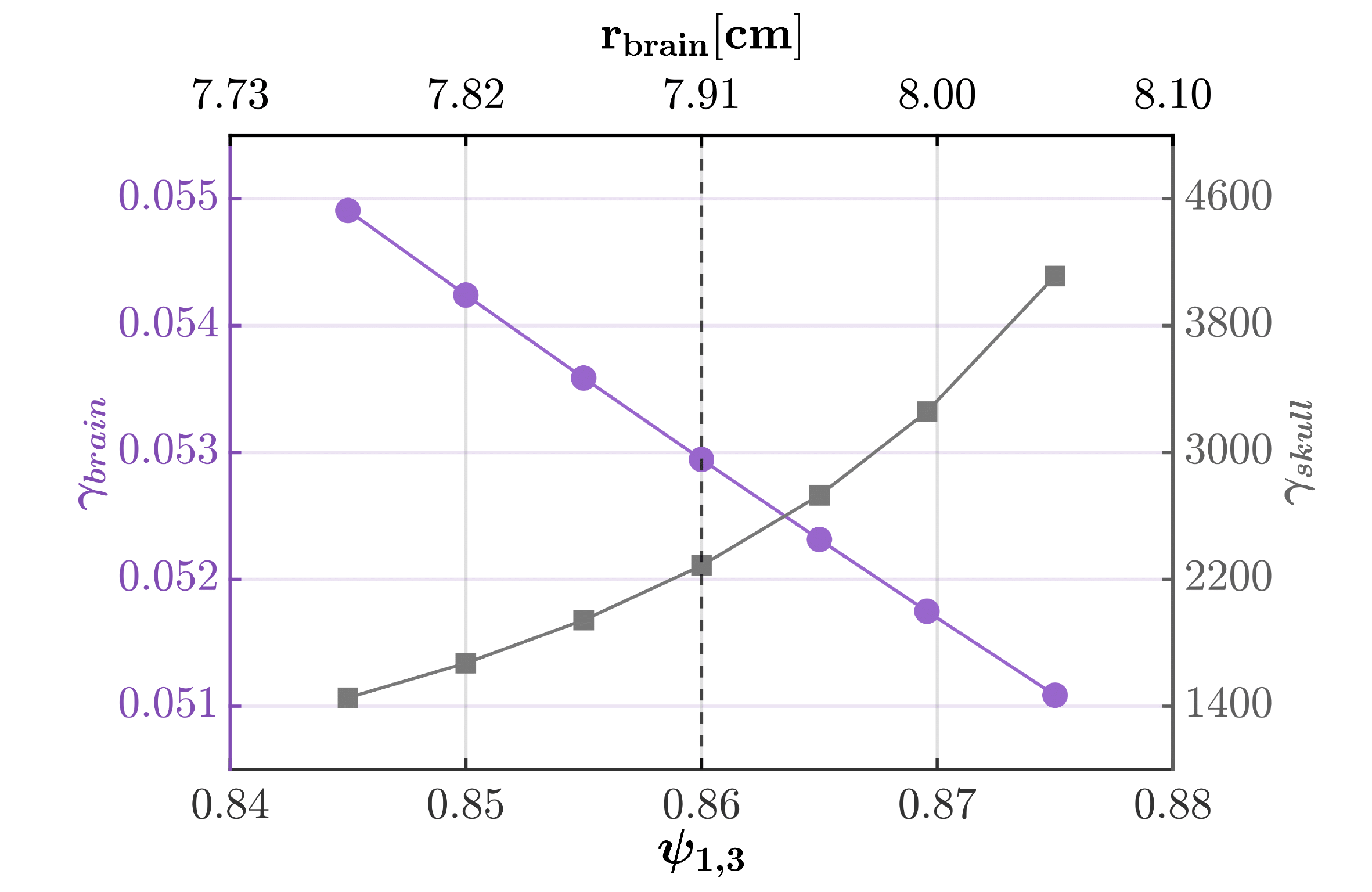}
    \caption{Polynomial fit of the $\gamma$ parameters for the brain and skull layers as a function of the brain-to-scalp radius ratio $\psi_{1,3}$. The upper axis also shows the corresponding brain radius $r_{\mathrm{brain}}$. The skull radius $r_{\mathrm{skull}}$ and scalp radius $r_{\mathrm{scalp}}$ are kept fixed. The dashed vertical line indicates the \(\gamma\) values associated with standard brain radius. The curves represent first-order polynomial fit for $\gamma_{\mathrm{brain}}$ (RMSE = $2.17\times10^{-3}$) and second-order polynomial fit for $\gamma_{\mathrm{skull}}$ (RMSE = $1.81\times10^{-3}$).
}
    \label{fig:bfit}
\end{figure}

\begin{figure}
    \centering
    \includegraphics[width=1\linewidth]{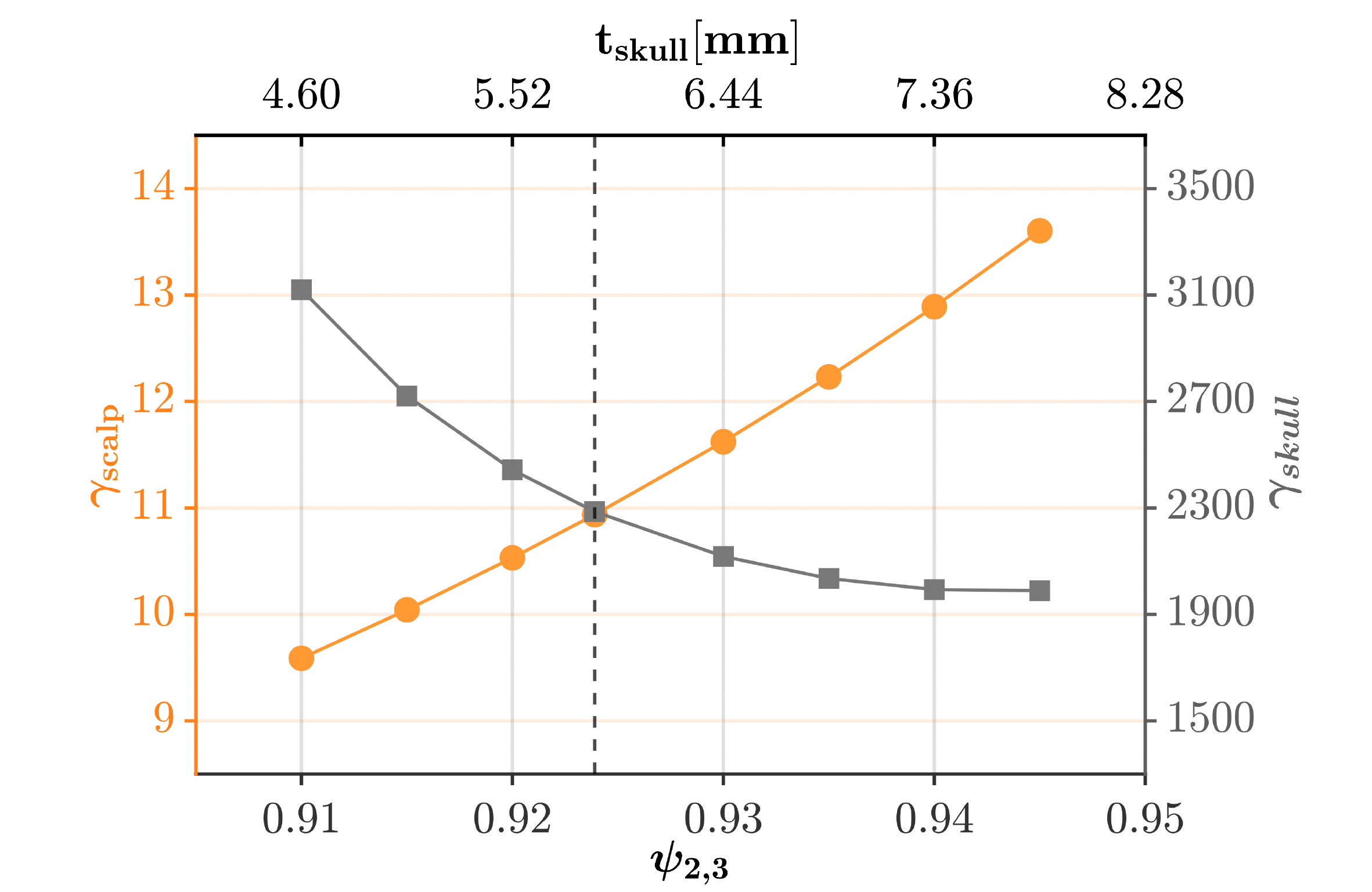}
    \caption{Polynomial fit of the $\gamma$ parameters for the skull and scalp layers as a function of the skull-to-scalp radius ratio, $\psi_{2,3}$. The upper axis also shows the corresponding skull thickness $t_{\mathrm{skull}}$. The brain radius $r_{\mathrm{brain}}$ and scalp radius $r_{\mathrm{scalp}}$ are kept fixed. The dashed vertical line indicates the \(\gamma\) values associated with standard skull thickness. The curves represent first-order polynomial fit for $\gamma_{\mathrm{scalp}}$ (RMSE = $5.39\times10^{-3}$) and second-order polynomial fit for $\gamma_{\mathrm{skull}}$ (RMSE = $4.81\times10^{-3}$).
}
    \label{fig:cfit}
\end{figure}

The first sweep analyzed the brain-to-scalp radius ratio, $\psi_{1,3} = r_{\mathrm{brain}} / r_{\mathrm{scalp}}$, varied from 0.845 to 0.875.  
The second sweep investigated the skull-to-scalp radius ratio, $\psi_{2,3} = r_{\mathrm{skull}} / r_{\mathrm{scalp}}$, which was varied between 0.910 and 0.945.  
These ranges correspond respectively to a $2.76\,\mathrm{mm}$ and $3.2\,\mathrm{mm}$ displacement of the brain-skull and skull-scalp interface radii, equivalent to an approximate $\pm20\%$ variation in skull thickness, 
the most influential yet also the most uncertain anatomical parameter affecting scalp potential estimation~\cite{chauveau2004effects,antonakakis2020inter}.  
For both ratios, the $\gamma_i$ coefficients curves were obtained through first- and second-order polynomial fits of numerically optimized data based on the semi-analytical reference solution (Figs.~\ref{fig:bfit} and~\ref{fig:cfit}).  

Finally, variations in overall head size were modeled by rescaling all impedance elements by a factor of $1/r_{\mathrm{scalp}}^2$.
 \vspace{3pt}
 
In the following section, the model is validated in an extended setting that accounts for dispersive behavior, with frequency-dependent electrical parameters, as well as concurrent geometric variations, to assess its ability to generalize beyond the baseline configuration.
In addition, the impact of neglecting displacement currents and dispersion effects is quantified to evaluate the errors introduced by such simplifications.
\section{Numerical Results}
\label{sec:results}
The accuracy of the proposed lumped model was evaluated against the semi-analytical formulation under dispersive conditions, using the \textit{in vivo} frequency-dependent conductivity and permittivity data reported by Wagner \textit{et al.}~\cite{wagner2014impact}. The comparison was carried out over 75 frequency points, logarithmically spaced between 10~Hz and 50~kHz.

Five dipole eccentricity values were considered ($\eta = 0.233$, $0.465$, $0.814$, $0.935$, and $0.966$), corresponding to dipole positions $r_{\mathrm{dip}} = 1.84$, $3.68$, $6.44$, $7.40$, and $7.64~\mathrm{cm}$, respectively. All dipoles were radially oriented and assigned a dipole moment of $p_r = 15~\mathrm{nA{\cdot}m}$. 
The skull thickness was varied between $4.6$ and $8.2~\mathrm{mm}$, while the brain and scalp radii were kept constant.
Finally, the influence of omitting dispersion and capacitive branches was analyzed.
\subsection{Circuit Model Validation}

\begin{figure}[t]
    \centering
    \includegraphics[width=1\linewidth]{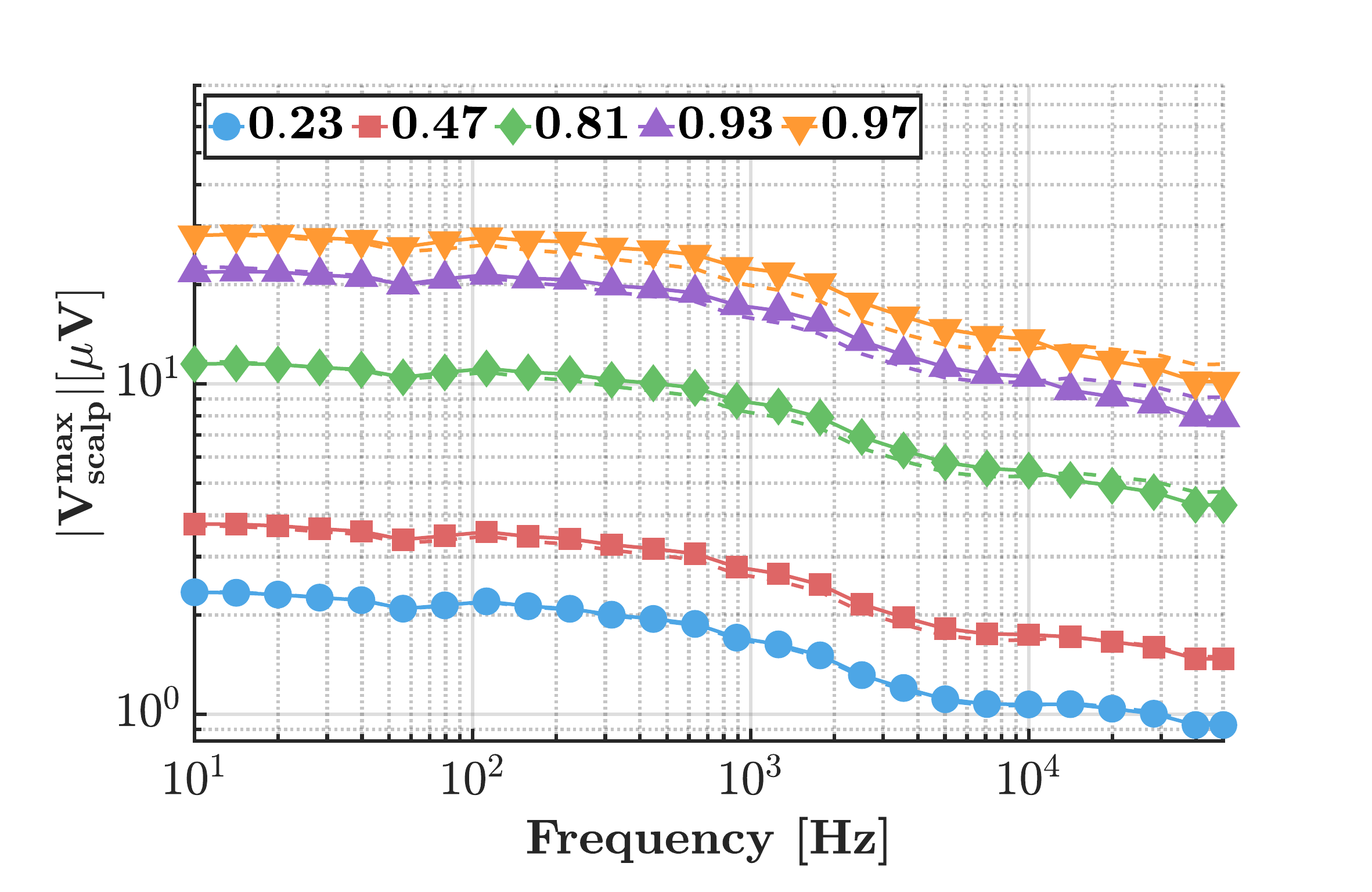}
    \caption{Frequency-domain comparison of the maximum scalp potential magnitude predicted by the semi-analytical model (solid lines) and the lumped RC circuit (dashed lines) for different dipole eccentricities.}
    \label{fig:error_ecc_plot}
\end{figure}

Fig.~\ref{fig:error_ecc_plot} shows the frequency response magnitude of the maximum scalp potential predicted for different dipole eccentricities.
Solid and dashed lines represent the semi-analytical reference and the circuit model estimates, respectively.
 
Minor deviations appear at high eccentricities, where the dipole approaches the brain-skull interface and local geometric inhomogeneities become more influential, as expected given the compact number of impedance blocks in the circuit representation.

\begin{figure}[t]
    \centering
    \includegraphics[width=1\linewidth]{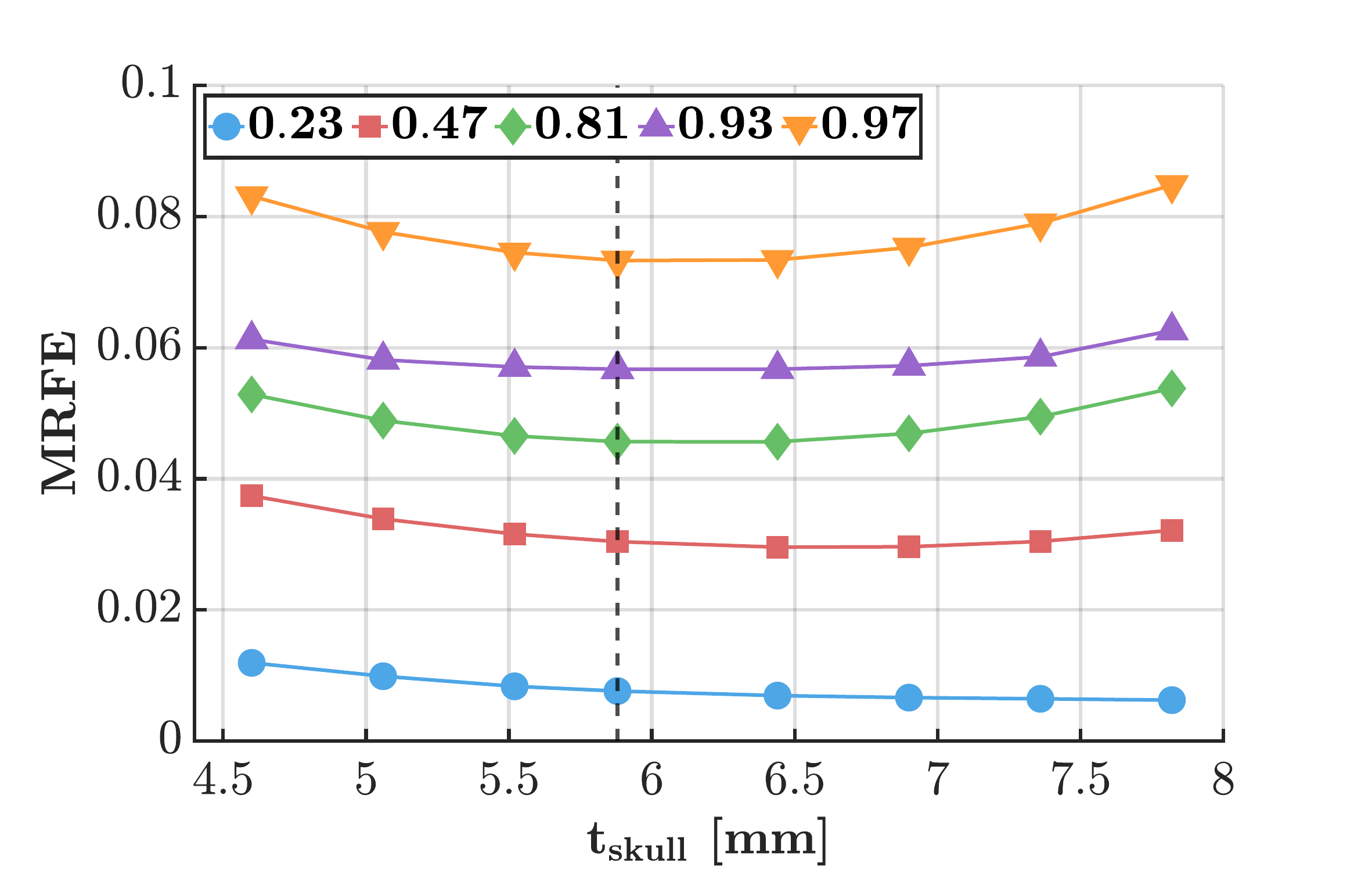}
    \caption{Mean Relative Frequency Error (MRFE) in \(|V_{\mathrm{scalp}}^{\mathrm{max}}|\) between the lumped model and the semi-analytical reference, evaluated over dipole eccentricity $\eta$ and skull thickness $t_{\mathrm{skull}}$. The dashed vertical line indicates the performances achieved with standard skull thickness.}
    \label{fig:mrfe_plot}
\end{figure}

The validation was further extended to assess the influence of skull thickness. 
Fig.~\ref{fig:mrfe_plot} reports the Mean Relative Frequency Error (MRFE) between the lumped model and the semi-analytical solution, evaluated over the 10~Hz--50~kHz frequency range as a function of both skull thickness and dipole eccentricity.
The MRFE is defined as
\begin{equation}
\mathrm{MRFE} = \frac{1}{N_f} \sum_{f}
\left|
\frac{V_{\mathrm{scalp,f}}^{\mathrm{Circuit}} - V_{\mathrm{scalp,f}}^{\mathrm{SSH}}}
{V_{\mathrm{scalp,f}}^{\mathrm{SSH}}}
\right|,
\end{equation}
where $N_f=75$ is the number of frequency samples considered, 
$V_{\mathrm{scalp}}^{\mathrm{Circuit}}$ denotes the scalp potential obtained from the proposed circuit model, 
and $V_{\mathrm{scalp}}^{\mathrm{SSH}}$ is the semi-analytical reference value from the spherical-harmonics formulation.

The MRFE increased moderately when the skull-scalp interface shifted closer to either the brain or the outer air boundary, with the strongest impact observed for high-eccentricity configurations.
Individually, skull thickness had a smaller effect compared to the error introduced by source eccentricity.

\subsection{Impact of Dispersion and Displacement Currents}

\begin{figure}[t]
    \centering
    \includegraphics[width=1\linewidth]{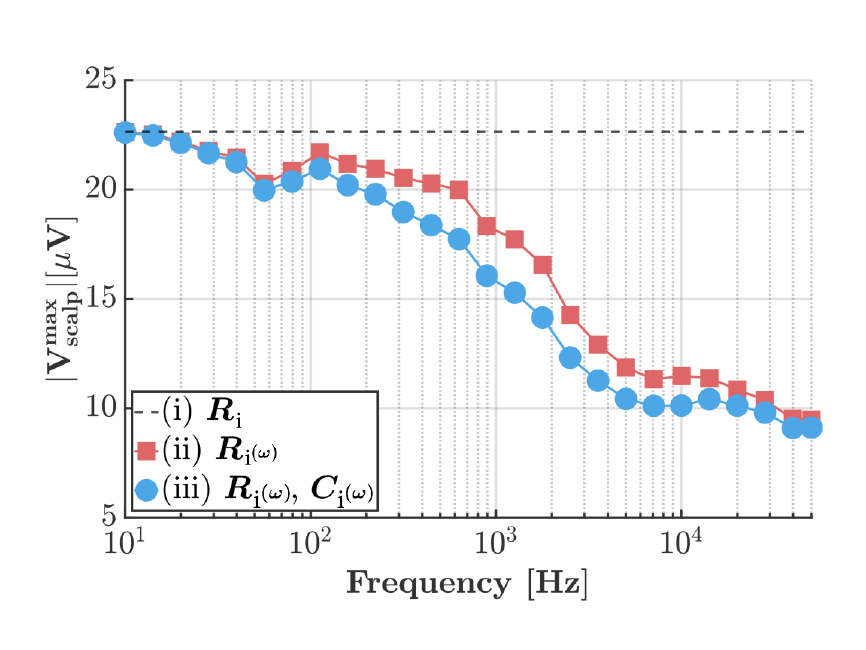}
    \caption{Frequency response of the maximum scalp potential magnitude $|V_{\mathrm{scalp}}^{\mathrm{max}}|$ predicted by the lumped circuit for a radial dipole at $r_{\mathrm{dip}}=7.40~\mathrm{cm}$ in the three-shell head (cf. Fig.~\ref{fig:headmodel}). The three cases correspond to: (i) ohmic, nondispersive $R_i$; (ii) dispersive $R_i(\omega)$ without displacement currents; and (iii) dispersive $R_i(\omega)$ and dispersive $C_i(\omega)$.}
    \label{fig:analstudy}
\end{figure}

The influence of frequency dispersion and displacement currents on scalp potential estimation was assessed using the proposed circuit model. 
Fig.~\ref{fig:analstudy} reports the frequency response of the maximum scalp potential magnitude, $|V_{\text{scalp}}^{\max}|$, for a three-shell spherical head with a radial brain dipole located $5~\mathrm{mm}$ beneath the brain-skull interface ($r_{\mathrm{dip}}=7.40~\mathrm{cm}$). 
Three configurations were analyzed:
\begin{itemize}
    \item (i) purely ohmic, neglecting both capacitances and resistive frequency dispersion ($R_i$ constant, $C_i = 0$);
    \item (ii) frequency-dispersive resistances without capacitances ($R_i(\omega)$, $C_i = 0$);
    \item (iii) frequency-dispersive resistances and capacitances ($R_i(\omega)$, $C_i(\omega)$).
\end{itemize}

The results showed that neglecting displacement currents or dispersion leads to substantial overestimation of $|V_{\text{scalp}}^{\max}|$. 
Case~(i) yields the largest deviation, with relative errors increasing already from $20~\mathrm{Hz}$ and reaching approximately $150\%$ at $50~\mathrm{kHz}$. 
Including frequency-dispersive conductivity alone [case~(ii)] reduces substantially the error; however, discrepancies exceeding $10\%$ still occur from around $450~\mathrm{Hz}$ up to about $17\%$ at $1.7~\mathrm{kHz}$. 
These findings confirm the necessity of including both displacement current and dispersive electrical parameters in head models to achieve accurate predictions in the $10~\mathrm{Hz}$--$50~\mathrm{kHz}$ range.


\section{Discussion and Conclusion}
\label{sec:discussion}

Validation results indicated a close agreement between the voltage potentials estimated by the proposed lumped-element model and the spherical-harmonics reference solution across the entire investigated frequency range, as well as under varying dipole eccentricities and skull thicknesses.
While the semi-analytical formulation remains valuable for validation, its applicability is confined to spherical geometries and it cannot be directly embedded in circuit-simulation frameworks. 
Conversely, the proposed equivalent circuit provides a compact electrical representation of the same physical behavior, enabling efficient simulation in circuit environments (e.g., SPICE) with a minimal number of elements. This offers a practical advantage over SSH models, which are generally incompatible with circuit netlists, and provides a foundation for future extensions to more general or realistic head geometries.

The observed dependence of the scalp potential error on skull thickness is consistent with the results reported by Chauveau \textit{et al.}~\cite{chauveau2004effects} for RMM formulations, while the increase in error with dipole eccentricity aligns with trends observed in both RMM, FEM and BEM forward models~\cite{medani2015fem,chauveau2004effects}. These outcomes confirm that the proposed simplified topology reproduces the main qualitative behaviors captured by more complex numerical approaches.

It is important to note, however, that these volumetric solvers substantially differ in scope. They are designed to evaluate potential distributions at multiple scalp locations, often at the expense of higher computational cost, limited suitability for circuit-level simulation, or a formulation restricted to static, resistive components. 
In contrast, the present model was conceived as a compact, frequency-aware equivalent circuit focused on estimating specific target potentials, thereby enabling faster simulation and easier integration into circuit-oriented design environments.

Although the proposed circuit employed a reduced number of impedance elements, it achieved accuracy levels in peak scalp potential estimation comparable to standard solvers, while maintaining substantially lower complexity. 
This makes it particularly suitable for rapid device testing, signal-to-noise ratio assessment, and real-time applications such as adaptive neuromodulation parameter tuning~\cite{fernandez2016transcranial,tian2024model}. 
Furthermore, the analysis of capacitive-branch and frequency dispersion contributions to the proposed circuit accuracy highlighted the importance of including both mechanisms in head models to ensure reliable predictions within the considered frequency range, in agreement with previous findings in the literature~\cite{gaugain2023quasi,wagner2014impact}.
 
At the current stage, validation has been limited to radially oriented dipole sources within concentric, isotropic layers. Future developments will address tangential sources, anisotropic conductivities, and more realistic anatomical configurations to further enhance the physiological accuracy of the head circuit model.

In summary, the proposed surrogate circuit framework provides an efficient approach for electrical head modeling.
By incorporating both dispersive and capacitive effects in a compact representation, it offers a useful tool for hardware system design in neural sensing and stimulation applications, with the potential to be extended to more complex geometries and physiological conditions in future work.
\section*{Acknowledgment}
This work was supported by the European Union through the CEREBRO project, which aims to develop the first EEG contrast medium for non-invasive, whole-brain functional imaging. It was also supported by the European Research Council (ERC) through the TurboEEG Proof of Concept grant.
\newpage
\bibliographystyle{ieeetr}       
\bibliography{references}        

%
\newpage
\begin{IEEEbiography}[{\includegraphics[width=1in,height=1.25in,clip,keepaspectratio]{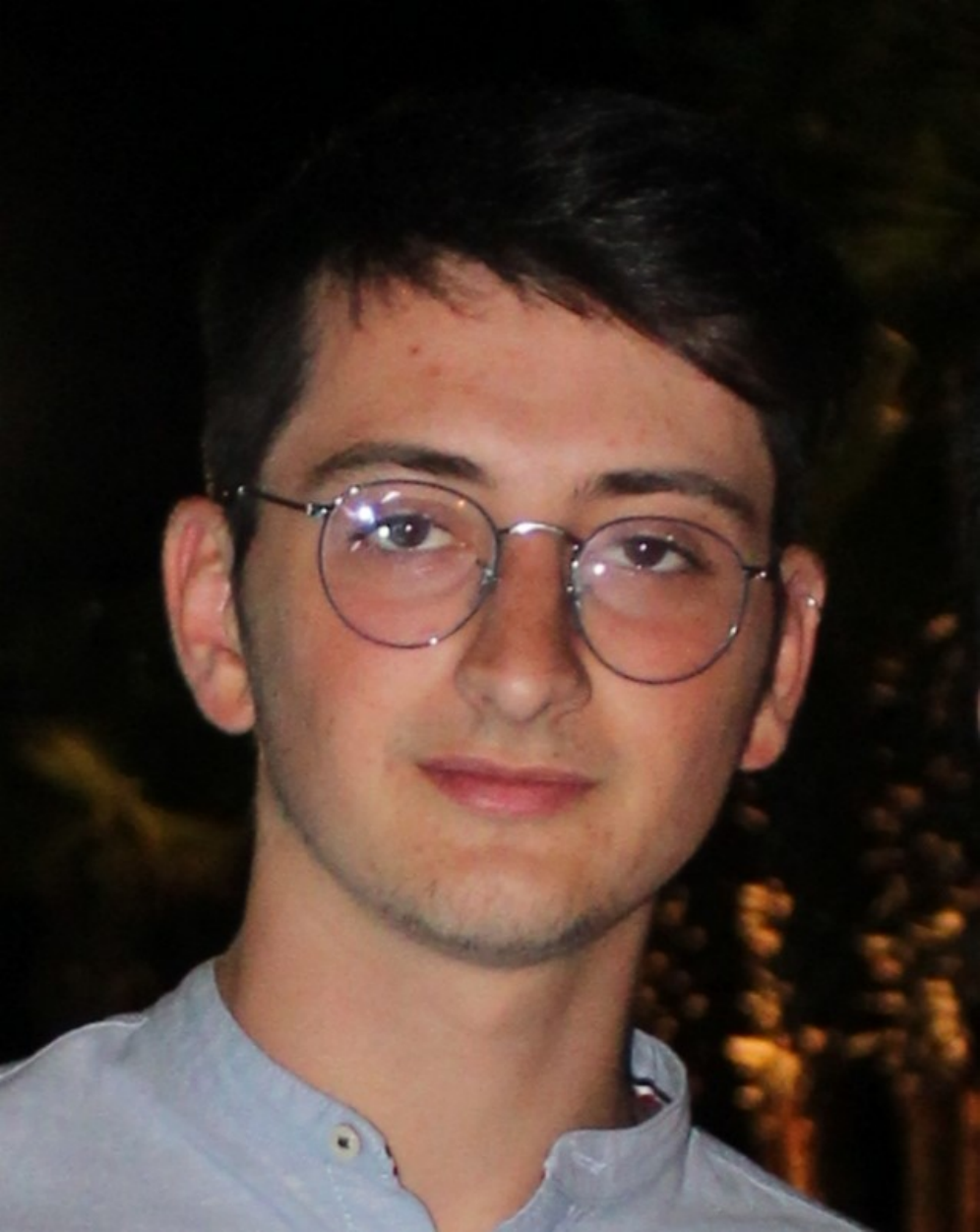}}]{Angelo Faccia}
(Graduate Student Member, IEEE) received the B.Sc. and M.Sc. degrees in biomedical engineering from Politecnico di Torino, Turin, Italy, in 2022 and 2024, respectively. He is currently pursuing the Ph.D. program in Electronics and Telecommunication engineering at the same institution, started in March 2025. His research interests include computational electromagnetics and brain imaging.
\end{IEEEbiography}

\begin{IEEEbiography}[{\includegraphics[width=1in,height=1.25in,clip,keepaspectratio]{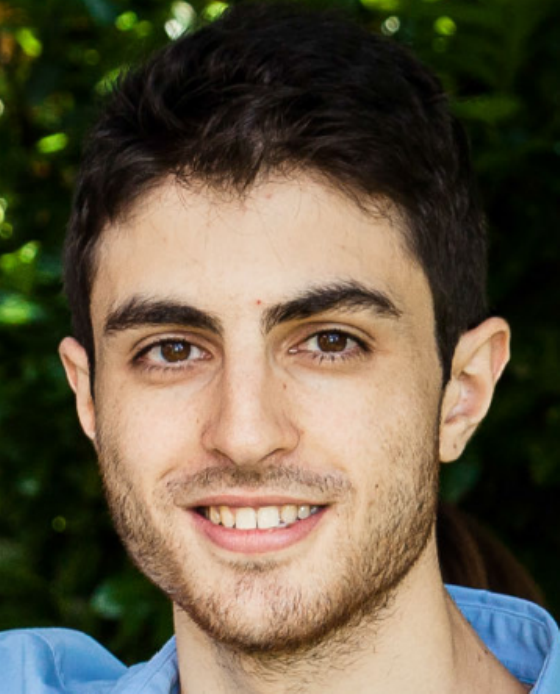}}]{Ermanno Citraro} received the B.Sc. and M.Sc. degrees in electrical engineering from the Politecnico di Torino, Turin, Italy, in 2017 and 2019, respectively, where he received his Ph.D. degree in 2025. After a two years' working experience in the semiconductor industry, he joined the Politecnico di Torino as a Research Associate in 2021. His research interests include computational electromagnetics, fast solvers, and brain imaging. Mr. Citraro has authored a paper that received
an honorable mention in International Union of Radio Science (URSI)/IEEE-APS 2022.
\end{IEEEbiography}

\begin{IEEEbiography}[{\includegraphics[width=1in,height=1.25in,clip,keepaspectratio]{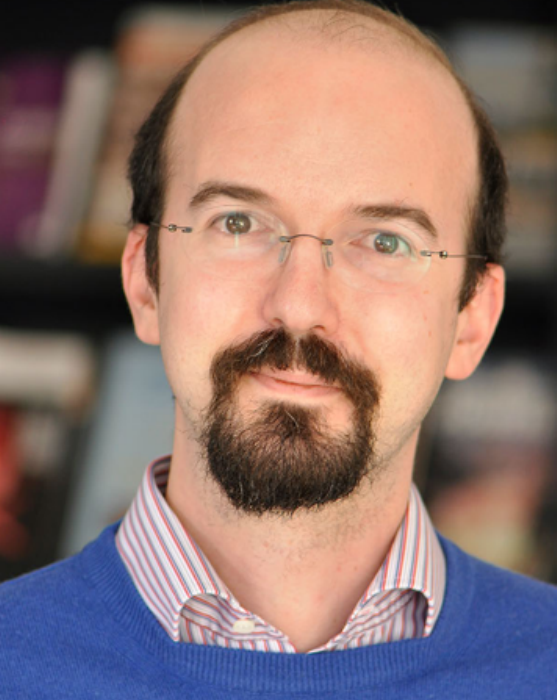}}]{Francesco P. Andriulli}
(Fellow, IEEE) received the Laurea degree in electrical engineering from the Politecnico di Torino, Turin, Italy, in 2004, the M.Sc.\ degree in electrical engineering and computer science from the University of Illinois at Chicago in 2004, and the Ph.D. in electrical engineering from the University of Michigan at Ann Arbor in 2008. From 2008 to 2010 he was a Research Associate with the Politecnico di Torino. From 2010 to 2017 he was an Associate Professor (2010--2014) and then a Full Professor with the \'Ecole Nationale Sup\'erieure Mines-T\'el\'ecom Atlantique (IMT Atlantique), Brest, France. Since 2017 he has been a Full Professor with the Politecnico di Torino. His research interests are in computational electromagnetics including frequency- and time-domain integral equation solvers, well-conditioned formulations, fast solvers, low-frequency electromagnetic analyses, and modeling techniques for antennas, wireless components, microwave circuits, and biomedical applications with a special focus on brain imaging.

Prof.~Andriulli received several best paper awards at conferences and symposia (URSI NA 2007, IEEE AP-S 2008, ICEAA IEEE-APWC 2015), also in coauthorship with his students and collaborators (EMTS 2025, ICEAA IEEE-APWC 2021, EMTS 2016, URSI-DE Meeting 2014, ICEAA 2009), with whom he also received a second prize conference paper (URSI GASS 2014), a third prize conference paper (IEEE AP-S 2018), seven honorable mention conference papers (ICEAA 2011, URSI/IEEE AP-S 2013, four in URSI/IEEE AP-S 2022, URSI/IEEE AP-S 2023), and three other finalist conference papers (URSI/IEEE AP-S 2012, URSI/IEEE AP-S 2007, URSI/IEEE AP-S 2006, URSI/IEEE AP-S 2022). Moreover, he received the 2014 IEEE AP-S Donald G.\ Dudley Jr.\ Undergraduate Teaching Award, the triennium 2014--2016 URSI Issac Koga Gold Medal, and the 2015 L.\ B.\ Felsen Award for Excellence in Electrodynamics.

Prof.~Andriulli is a Fellow of the IEEE and of the International Union of Radio Science (URSI), and a member of Eta Kappa Nu, Tau Beta Pi, and Phi Kappa Phi. He serves as the 2026 President-Elect of the IEEE Antennas and Propagation Society and served as IEEE AP-S Vice-President of Publications in 2025, as Editor-in-Chief of the \textsc{IEEE Antennas and Propagation Magazine}, as Track Editor for the \textsc{IEEE Transactions on Antennas and Propagation}, and as an Associate Editor for the \textsc{IEEE Antennas and Wireless Propagation Letters}, \textsc{IEEE Access}, \textsc{URSI Radio Science Letters}, and \textsc{IET-MAP}.
\end{IEEEbiography}





\end{document}